\documentclass[%
reprint,
 amsmath,amssymb,
 aps,
]{revtex4-2}

\usepackage{amsmath}
\usepackage{graphicx}
\usepackage{dcolumn}
\usepackage{bm}
\usepackage{amsmath}
\usepackage{float}
\usepackage{mathtools}  
\usepackage{xfrac}

 \usepackage{hyperref}
 \hypersetup{
     colorlinks=true,
     linkcolor=blue,
     filecolor=blue,
     citecolor = blue,      
     urlcolor=blue,
     }
     \usepackage{tikz,xcolor,hyperref}

\definecolor{lime}{HTML}{A6CE39}
\DeclareRobustCommand{\orcidicon}{%
	\begin{tikzpicture}
	\draw[lime, fill=lime] (0,0) 
	circle [radius=0.16] 
	node[white] {{\fontfamily{qag}\selectfont \tiny ID}};
	\draw[white, fill=white] (-0.0625,0.095) 
	circle [radius=0.007];
	\end{tikzpicture}
	\hspace{-3.5mm}
}

\foreach \x in {A, ..., Z}{%
	\expandafter\xdef\csname orcid\x\endcsname{\noexpand\href{https://orcid.org/\csname orcidauthor\x\endcsname}{\noexpand\orcidicon}}
}

\begin{document}

\preprint{APS/123-QED}

\title{Absorbance Enhancement of Monolayer MoS$_2$ in a Perfect Absorbing System}
\thanks{\# These authors contribute equally.}

\author{Xia Zhang\orcidA{}\textsuperscript{\#}, Julia Lawless\orcidC{}\textsuperscript{\#}, Jing Li, John F. Donegan\orcidD{}}
 \affiliation{%
School of Physics, CRANN and AMBER, Trinity College Dublin, Dublin, Ireland
}
\author{Lisanne Peters\orcidF{}}
 \affiliation{School of Chemistry and AMBER, Trinity College Dublin,  Dublin, Ireland}
\author{Niall McEvoy\orcidE{}}
\affiliation{School of Chemistry and AMBER, Trinity College Dublin, Dublin, Ireland}

\author{A. Louise Bradley\orcidB{}}%
\email{bradlel@tcd.ie}
 \affiliation{%
School of Physics, CRANN and AMBER, Trinity College Dublin, Dublin, Ireland
}%


\begin{abstract}
We reveal numerically and experimentally that dielectric resonances can enhance the absorbance and emission of monolayer MoS$_2$. By quantifying the absorbance of the Si disk resonators and the monolayer MoS$_2$ separately, a model taking into account the absorbance as well as quantum efficiency modifications by the dielectric disk resonators successfully explains the observed emission enhancement under the normal light incidence. It is demonstrated that the experimentally observed emission enhancement at different pump wavelength results from the absorbance enhancement, which compensates the emission quenching by the disk resonators. In order to further maximize the absorbance value of monolayer MoS$_2$, a perfect absorbing structure is proposed. By placing a Au mirror beneath the Si nanodisks, the incident electromagnetic power is fully absorbed by the hybrid monolayer MoS$_2$-disk system. It is demonstrated that the electromagnetic power is re-distributed within the hybrid structure and 53\% of the total power is absorbed by the monolayer MoS$_2$ at the perfect absorbing wavelength. 
\end{abstract}

\maketitle
\section{Introduction}
Tailoring the absorbance and emission properties of an emitter is of great interest in terms of both understanding the fundamental physics and the application prospects \cite{novotny2012principles}. Both are determined by the electromagnetic environment of the emitter, which can be realized by positioning the emitter in the near field of a resonant cavity, such as a photonic-crystal nanocavity \cite{ogawa2004control,wang2016giant, aoki2008coupling}, plasmonic nanoantenna \cite{kuhn2006enhancement, koenderink2017single} or grating \cite{zhao2015resonance,le2019giant}. 
Particularly, both
properties are fundamentally correlated. The Purcell effect describes the modification of the spontaneous emission rate due to the emitters environment through the refractive index, and the resonator mode volume and quality factor \cite{purcell1946resonance}. The spontaneous emission can also be modified by the absorbance enhancement. Such absorbance and emission tailoring belong to the regime of weak light-matter coupling or perturbative regime. Monolayer two-dimentional (2D) transition metal dichalcognides (TMDCs) have become core materials for nanoscale devices due to reduced dimensionality, ease of integration and advanced optical and electronic functionalities, such as nanolaser \cite{khajavikhan2012thresholdless}, photodetector \cite{xia2009ultrafast} and nonlinear optic elements \cite{jiang2018gate}.

Monolayer TMDCs exhibits a direct bandgap in the visible and near-infrared wavelength \cite{mak2010atomically}. However, due to the atomic thickness of the layer and following the Beer-Lambert absorbance law, low optical absorbance and emission of monolayers TMDCs poses an obstacle for conversion of photon energy into other forms of energy. Intensive efforts have therefore been devoted to enhance the absorbance or emission of monolayer TMDCs, such as by employing a photonic crystal slab backed by a perfect electric conductor mirror \cite{piper2014total,piper2014total}, a multilayer photonic structure including a dielectric grating, spacer and a metal film \cite{lu2017nearly} and a sandwiched photonic crystal slab with air holes and silver mirror \cite{li2017total} or through a metasurface \cite{butun2015enhanced}, photonic hypercrystals \cite{galfsky2016broadband}, a photonic Fano resonance \cite{zhang2017unidirectional} and a plasmonic antenna \cite{palacios2017identifying, Butun2017}.
The maximum absorbance that a structure can achieve is 1, referred to as a perfect absorber or coherent perfect absorption (CPA) \cite{chong2010coherent,baranov2017coherent}. CPA can involve interference due to two or more input waves including counter-propagating input waves, and can be manipulated by varying the relative phase of the two inputs \cite{baranov2017coherent}. Due to its versatile manipulation, such configurations are useful for applications in modulators and optical switches.

\begin{figure}[H]
\includegraphics[width=1\linewidth]{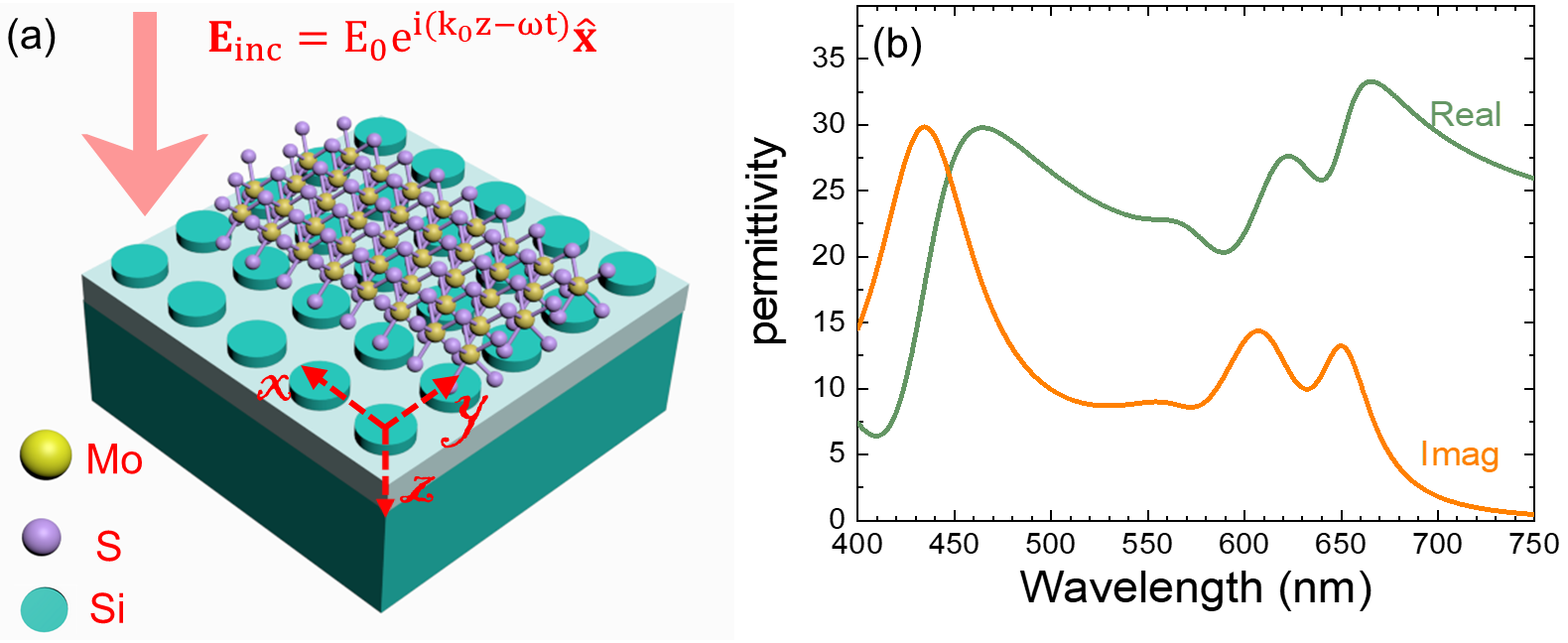}
\caption{\label{fig:sch} (a) Schematic of the hybrid Si nanodisk metasurface and monolayer MoS$_2$ on SiO$_2$/Si substrate. A normally incident plane wave propagates along $z$ direction and polarizes along $x$ direction, $\rm  {\textbf{E}_{inc}}= E_0 e^{i(k_0z-\omega t)}\hat{\textbf{x}}$. (b) The real and imaginary components of the dielectric permittivity of monolayer MoS$_2$. }
\end{figure}

In this work, we demonstrate experimentally that Si nanodisk resonators are capable of confining the energy within monolayer MoS$_2$ to achieve absorbance enhancement. The signature of absorbance enhancement via increased emission is experimentally observed. The emission enhancement as well as the absorbance of monolayer MoS$_2$ are quantified, and agree with the experimental observation. Based on this, we further explore the use of a mirror to increase the absorbance of monolayer MoS$_2$, wherein a perfect absorbing system is proposed enabling more than 50\% of the total power to be absorbed by the monolayer MoS$_2$.

\section{Experiment and Model}

\begin{figure}[htbp]
\includegraphics[width=1\linewidth]{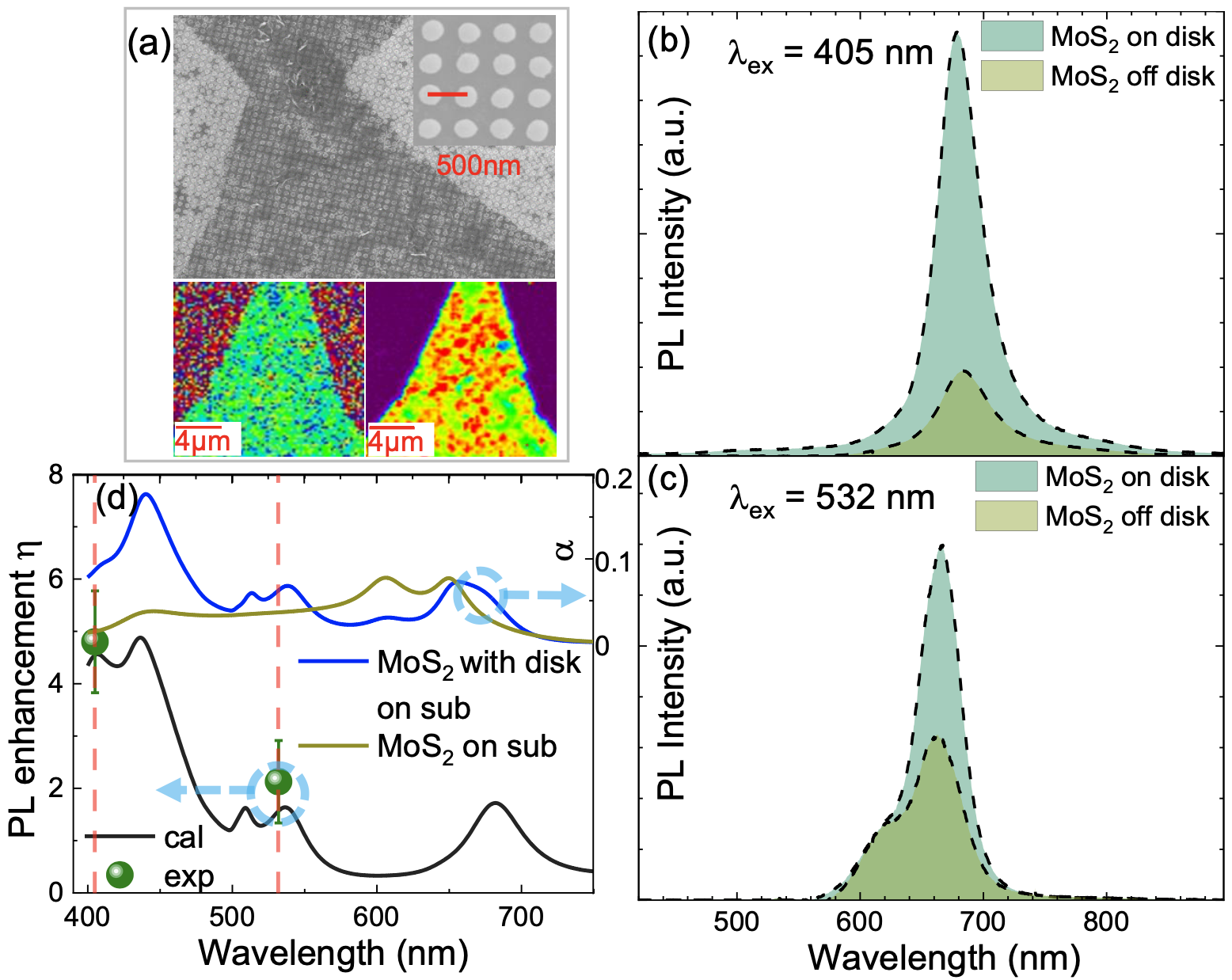}
\caption{\label{fig:exp} (a) The scanning electron microscope images (top), Raman (bottom left) and PL (bottom right) maps of the fabricated hybrid MoS$_2$-Si nanodisk sample on SiO$_2$/Si substrate for radius $r$ = 160 nm. Experimentally measured PL spectra of monolayer MoS$_2$, on disk and off disk, with the excitation wavelengths of (b) $\lambda_{ex} = 405$ nm and (c) $\lambda_{ex} = 532 $ nm, respectively, denoted as vertical dashed lines. (d) Top right: The calculated absorbance, $\rm \alpha$ of monolayer MoS$_2$ on a SiO$_2$/Si substrate and that hybridized with Si disks on a SiO$_2$/Si substrate. Bottom left: The experimentally measured and calculated PL enhancement $\rm \eta$. }
\end{figure}

\begin{figure*}[htbp]
\includegraphics[width=0.9\linewidth]{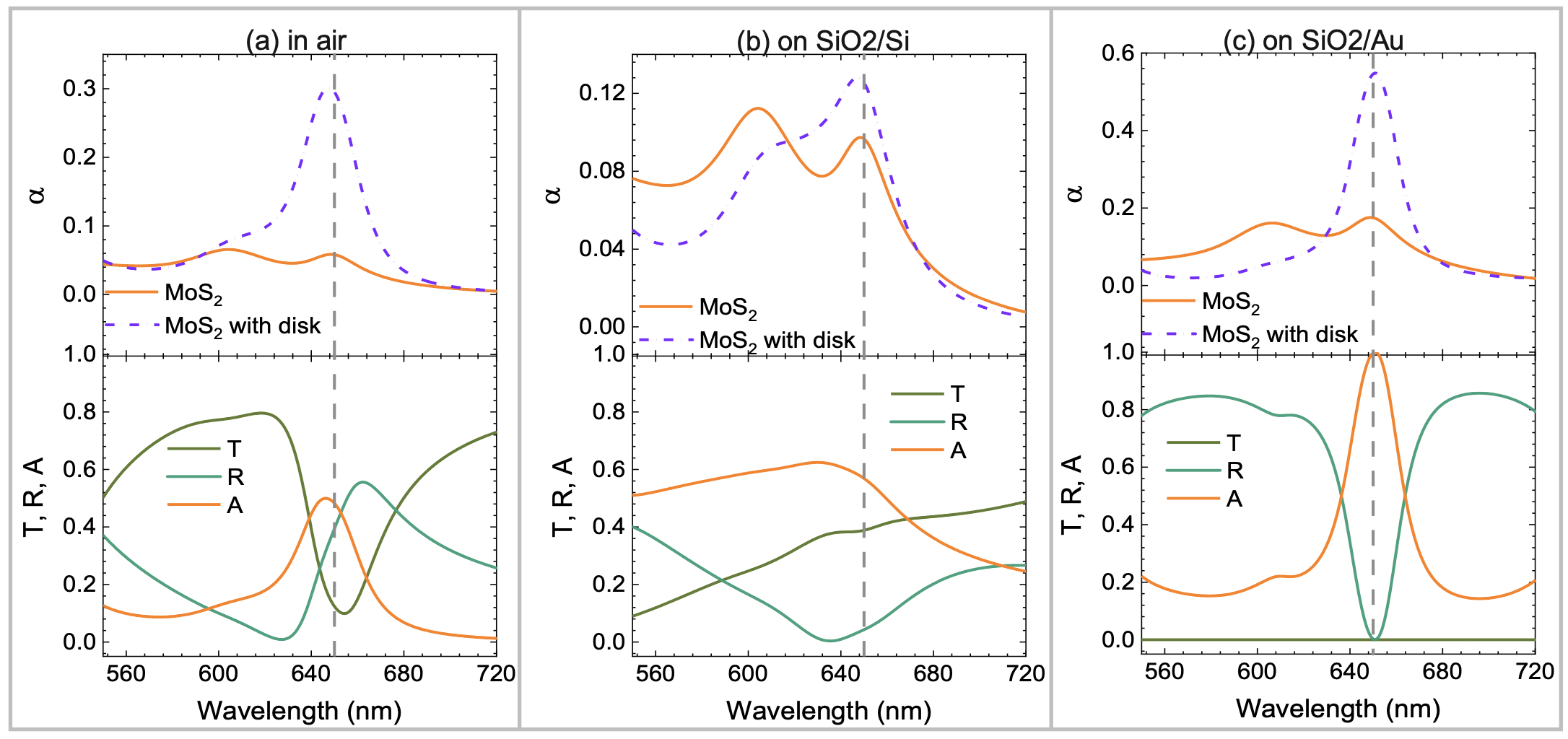}
\caption{\label{fig:abs} The separate MoS$_2$ monolayer and the hybrid MoS$_2$ monolayer-disk structures are examined for three cases: (a) in air, (b) on SiO$_2$/Si substrate and (c) on SiO$_2$/Au substrate, respectively. The disk has the radius, $\rm r$ = 150 nm; the height, $\rm h$ = 45 nm and the spacer thickness, $\rm s$ = 100 nm. Top panel: the absorbance of the MoS$_2$ monolayer in the alone and hybrid cases. Bottom panel: The reflectance (R), transmittance (T) and absorbance (A) of the total structure. }
\end{figure*}

A schematic of the hybrid emitter-resonator structure can be seen in Fig.~\ref{fig:sch}, which shows the MoS$_2$ monolayer on the Si nanodisk resonators, sitting on a SiO$_2$/Si substrate. The permittivity of monolayer MoS$_2$ is take from Ref. \cite{li2014measurement} and is shown in Fig.~\ref{fig:sch} (b). The Si nanodisk metasurface is fabricated on a Si-on-insulator (SOI) wafer (45 nm top Si thickness, 150 nm buried oxide thickness, Soitec). Electron-beam lithography is performed on spin-coated PMMA 950 layer (A3, 3000 rpm) by Electron beam (Elionix ELS 7700), followed by the development in MIBK:IPA 1:3. A 20 nm thickness of chromium (Cr) layer was deposited by e-beam evaporator (Temescal). After lift-off process in hot Remover 1165, the Cr layer was used as a hard mask, the pattern was transferred to the SOI substrate by Inductively Coupled Plasma Etching (ICP) through the top Si layer with an over-etched thickness of (10$\pm$3) nm BOX layer, verified by ellipsometry. The thickness of SiO$_2$ is 140 nm. The Cr mask was finally removed using a commercial Cr etchant from Sigma. 
The scanning electron microscope images, Raman and photoluminescence (PL) maps of the fabricated hybrid sample are shown in Fig.~\ref{fig:exp} (a) for the disk radius $r$ = 160 nm and the period, $p_x=p_y$ = 500 nm. The experimental measurement details, including reflectance spectra and Raman spectra proving the monolayer nature of MoS$_2$ can be found in the Supplemental Material, with further details in Ref. \cite{2021PRMsupp}. As can be seen in Fig.~\ref{fig:exp} (b) and (c), the measured emission intensity displays an enhancement due to the presence of the nanodisk resonators at both excitation wavelengths of $\rm \lambda_{ex} = 405$ nm and $\rm \lambda_{ex} = 532$ nm. The PL enhancements defined as $\eta$, at both pump wavelengths, denoted as grey dash lines, are shown in Fig.~\ref{fig:exp} (d).

\begin{figure*}[htbp]
\includegraphics[width=0.9\linewidth]{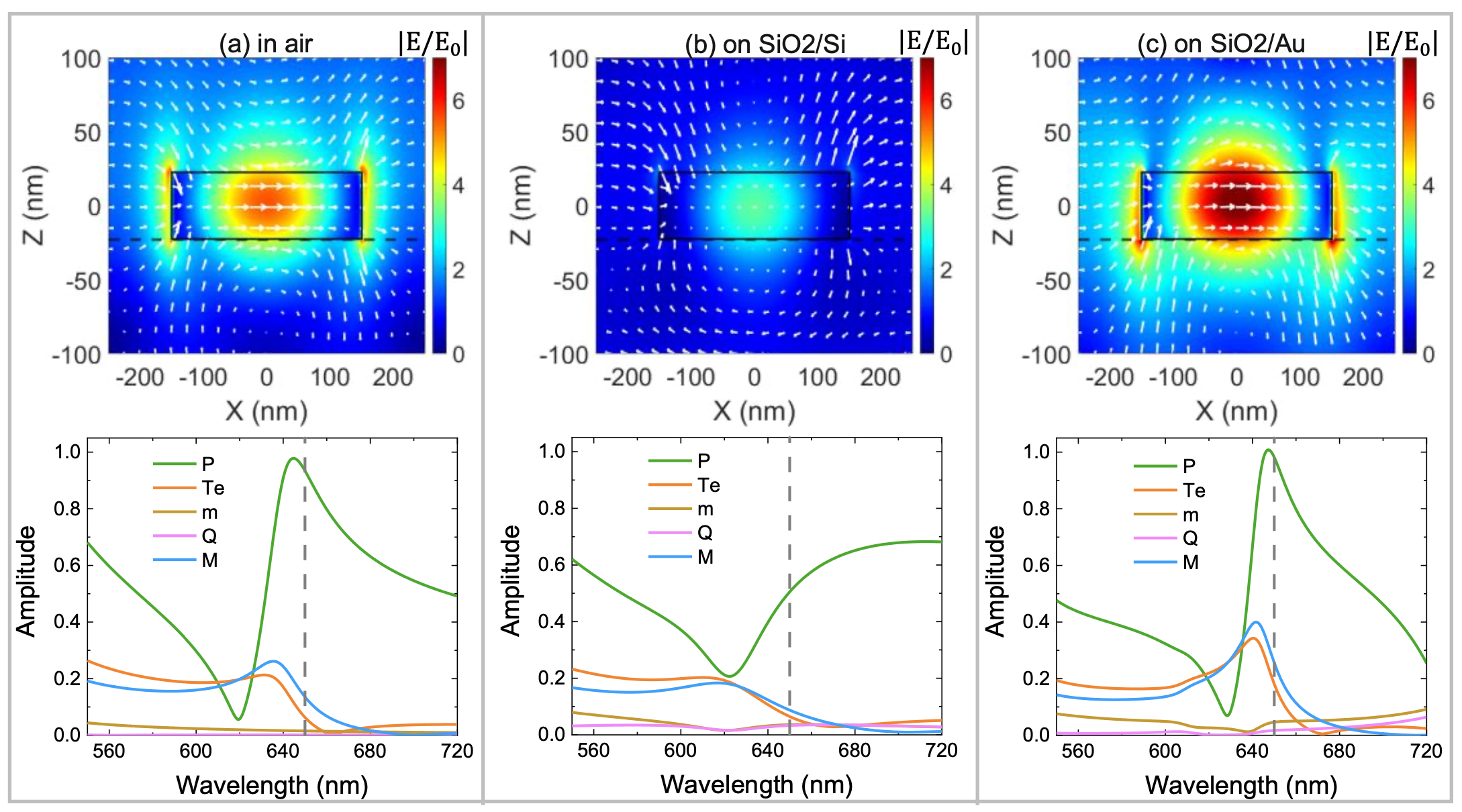}
\caption{\label{fig:field} Top panel: The simulated amplitude ratio of electric field, $|\mathbf{E}/\mathbf{E}_0|$ in $x-z$ plane at $y$ = 0 at the inspected absorbance wavelength of 646 nm. The disk has the radius, $\rm r$ = 150 nm, the height, $\rm h$ = 45 nm and the spacer thickness of $\rm s$ = 100 nm. The black dash line illustrates where the MoS$_2$ monolayer is and the incident light propagates along $+z$ direction.  Bottom panel: The amplitude of the multipolar modes for reflectance and transmittance coefficients: including the electric dipole $\rm \textbf{P}$, electric toroidal dipole $\rm \textbf{Te}$, magnetic dipole $\rm \textbf{m}$, electric quadrupole $\rm \textbf{Q}$ and magnetic quadrupole $\rm \textbf{M}$.}
\end{figure*}

In order to investigate the physics driving the PL enhancement, a model linking absorbance and emission is required. Following the law of power conservation, the sum of reflectance, transmittance and absorbance is unity. Therefore, the absorbance, $\rm A$ is typically determined as \cite{Butun2017}
\begin{equation}
\label{Eq:absRT}
 \rm A = 1 - R - T
\end{equation}
where $\rm R$ and $\rm T$ represent the reflectance and transmittance respectively. Eq.~\ref{Eq:absRT} offers a flexible way, numerically and experimentally, to quantify the absorbance value of the hybrid or the alone structures. However, in the hybrid sample aiming for the quantification of the absorbance of constituting elements, such as the monolayer MoS$_2$ or the Si nanodisk, it is critical to separate the power dissipation within the disk and monolayer MoS$_2$. The absorbance of a nanostructure, here the monolayer MoS$_2$ or the disk metasurface, is the ratio of the
total absorbed power within a volume $\rm V$ ($\rm P_{abs}$) to the incoming power over the exposed surface area $\rm S_0$ \cite{jackson1999classical,Baffou2009}, the absorbance under normal incidence can also be calculated analytically as 
\begin{equation}
\label{Eq:abspower}
\rm\alpha=\frac{P_{abs}}{P_{inc}}=\frac{k_0}{S_0|E_0|^2}\iiint_VIm[\epsilon (\textbf{r})]|\textbf{E}(\textbf{r})|^2 dV
\end{equation}
where $\rm P_{abs}=\frac{1}{2}c\epsilon_0 k_0\iiint_VIm(\epsilon)|\textbf{E}(\textbf{r})|^2 dV$\cite{zhao2014enhancement,lu2017nearly}. $|\rm \textbf{E}(\textbf{r})|$ is the amplitude of the electric field within the layer of monolayer MoS$_2$ or Si disk resonators. $\epsilon$ is the corresponding dielectric permittivity. The integration is performed in one unit cell over the volume $\rm V$ and $\rm V = p_x\times p_y\times h (t_0)$, where $\rm h$ and $\rm t_0$ denote the height of the disk and the thickness of the MoS$_2$ monolayer respectively. 
$\rm P_{inc}$ denotes the incident source power calculated over one unit cell in free-space, $\rm P_{inc} = \frac{1}{2}c\epsilon_0S_0 |{E}_0|^2$. $k_0$ is the free-space wavevector and $c$ is the free-space light speed. $\rm S_0$ is the area of one unit cell, $\rm S_0=p_x\times p_y$. $|\rm {E}_0|$ is the electric field amplitude of the incident light, which is $\rm  {\textbf{E}_{inc}}= E_0 e^{i(k_0z-\omega t)}\hat{\textbf{x}}$. Eq.~\ref{Eq:absRT} is used to calculate $\rm A$, the absorbance of a structure, which can be either the complete hybrid structure or the bare metasurface or the monolayer in the absence of the metasurface. Whereas $\alpha$ is used to quantify the absorbance of each component within the complete hybrid structure. A detailed discussion on the absorbance determination for periodic structures can be seen elsewhere \cite{brenner2010aspects}.

As can be seen in Fig.~\ref{fig:exp} (d) upper panel, the absorbance of monolayer MoS$_2$, $\rm \alpha$ is enhanced due to the influence of disk resonators at 405 nm and 532 nm, indicated as vertical dash lines.  The absorbance enhancement at different pump wavelengths is quantified as the ratio of the absorbance of monolayer MoS$_2$ on disk to that off disk as $\rm \alpha(\lambda_{pump})/\alpha^{0}(\lambda_{pump})$. Furthermore, as well as the effect of Si nanodisk on the absorbance,  the effect on the quantum efficiency need to be considered. The MoS$_2$ monolayer can be represented as an array of in-plane dipoles, the decay rates can be described by a simplified two-level quantum approach. The excitonic population relaxation states, $\gamma_{ex}$, includes the radiative part $\gamma^{r}_{ex}$ and nonradiative part, $\gamma^{nr}_{ex}$ as $\rm \gamma_{ex}=\gamma^{r}_{ex}+\gamma^{nr}_{ex}$. The quantum efficiency of the quantum system, is $\rm q=\gamma^{r}_{ex}/(\gamma^{r}_{ex}+\gamma^{nr}_{ex})$. The effective PL enhancement factor, $\eta$ is quantified as \cite{koenderink2017single,Sortino2019}

\begin{equation}
\label{PL}
  \rm \eta= \frac{\alpha(\lambda_{pump})}{{\alpha}^{0}(\lambda_{pump})}\cdot\frac{q(\lambda_{em})}{q^0(\lambda_{em})} 
\end{equation}
where $\rm \alpha_(\lambda_{pump})$ and $\rm \alpha^{0}(\lambda_{pump})$ represent the absorbance of the MoS$_2$ monolayer at the pump wavelength on disk, and off disk, respectively. $\rm q(\lambda_{em})$ and $\rm q^0(\lambda_{em})$ represent the quantum efficiency at the peak emission wavelength of the MoS$_2$ monolayer in the hybrid structure on the Si disk, and the MoS$_2$ monolayer off disk, respectively. The quantum efficiency is calculated at the peak emission wavelength $\rm \lambda_{em}$ = 673 nm. $\rm \lambda_{pump}$ = 405 nm and 532 nm are used in the absorbance simulation corresponding to the experimental measurements.

All the calculations are performed by commercial finite-difference-time-domain (FDTD) software (Lumerical Solutions Inc.). In the calculation, including R, T, A and $\rm P_{abs}$, a linearly polarized, normally incident plane wave source is applied, $\rm  {\textbf{E}_{inc}}= E_0 e^{i(kz-\omega t)}\hat{\textbf{x}}$. Periodic boundary conditions are applied in the $x-y$ directions with the pitch $p_x = p_y = 500 $ nm. A perfectly matched layer boundary condition is employed in the $z$ direction. The monolayer MoS$_2$ is modeled as a thin sheet of thickness of 0.75 nm, with the permittivity of MoS$_2$ \cite{li2014measurement}. The numerically simulated reflectance spectra match with the experimentally measured curves, which can be found in the Supplementary Materials, with further information in Ref. \cite{2021PRMsupp}. For the quantum efficiency calculation, the monolayer MoS$_2$ can be treated as a point dipole emitter due to its subwavelength exciton coherence length and $\sim$ 1 nm exciton Bohr radius \cite{fogler2008effect, zhang2014absorption,zhang2017unidirectional}. The quantum efficiency calculation is averaged over 11 different positions relative to the nanodisk resonator within one unit cell. The detailed calculation can be seen in Ref. \cite{2021PRMsupp}. The origin of the coordinate vector is placed at the center of nanodisk. The dipole positions for simulations are: $x_0$ ranges from 0 to 500 nm 
in steps of 50 nm, $y_0$ = 0, $z_0 = (h+t_0)/2$ = 22.875 nm, where $h$ = 45 nm for the disk height and $t_0$ = 0.75 nm for the measured thickness of the monolayer MoS2.  The dipole is oriented parallel to the substrate surface. A value of 0.74 is obtained for the ratio $\rm q(\lambda_{em})/q^0(\lambda_{em})$, which indicates the substrate results in the emission quenching of the MoS$_2$ monolayer. However, the absorbance enhancement compensates the losses. As can be seen in Fig.~\ref{fig:exp} (d) for the disk radius of $r= $160 nm, the calculated PL enhancement $\eta$ following Eq.~\ref{PL} shows agreement with the experimentally measured data, which validates our model. It also demonstrates that the observed emission enhancement mainly arises from the absorbance enhancement.

\section{Perfect Absorber: Mirror Effect}

Absorbance enhancement has been realized using the Si disk resonators on SiO$_2$/Si substrate, however, the absorbance value is still quite low, especially at the A exciton peak. In this section, we aim to increase the absorbance value employing the substrate effect. For a clear physical picture of the effect of the substrate, the calculated absorbance spectra of the hybrid disk-MoS$_2$ monolayer embedded in air medium, sitting on SiO$_2$/Si substrate as well as on SiO$_2$/Au can be seen in Fig.~\ref{fig:abs}. The inspected wavelength at 646 nm is indicated by the vertical dashed line. It can be seen that the absorbance value of monolayer MoS$_2$ without disk resonators is below 10\% in all cases (5.7\% in air, 9.5\% on SiO$_2$/Si and 4.6\% on SiO$_2$/Au, respectively). Due to the resonance effect producing enhanced electric field with the nanodisks, the absorbance of monolayer MoS$_2$ is enhanced at 646 nm (30\% in air, 12.7\% on SiO$_2$/Si and 53\% on SiO$_2$/Au respectively).

Furthermore, the reflectance and transmittance as well as the absorbance spectra of the hybrid structures are also shown, where $\rm R$ and $\rm T$ are obtained by FDTD simulation and $\rm A$ is calculated by Eq.~\ref{Eq:absRT}. Clearly as seen from Fig.~\ref{fig:abs} (c), the whole structure becomes a perfect absorber at the inspected wavelength ($\rm A = 1$). Due to the mirror effect contributed by the gold plane, the transmittance becomes zero, Eq.~\ref{Eq:absRT} becomes $\rm A = 1 - R$. The electric field maps are shown at the inspected wavelength for all three cases, where the electromagnetic energy is confined inside or in the near field of the nanodisk resonator. 
The SiO$_2$/Si substrate yields the lowest field amplitude, which results from the loss due to the Si substrate. A much larger electric field amplitude is obtained for the hybrid structure on a SiO$_2$/Au substrate. In particular, at the position of the monolayer MoS$_2$ shown by the black dashed lines, obvious field concentration is seen compared with that suspended in air or on the SiO$_2$/Si substrate. This field concentration drives the large absorbance value of monolayer MoS$_2$. CPA has been observed in other systems where perfect absorbance is achieved when the scattered components interfere destructively \cite{chong2010coherent, piper2014total, baranov2017coherent}. Here in the proposed structure on the SiO$_2$/Au substrate, the gold mirror reflects the transmitted wave through the hybrid structure producing destructive interference in reflection and enhanced absorbance.

\begin{figure}[htbp]
\includegraphics[width=0.8\linewidth]{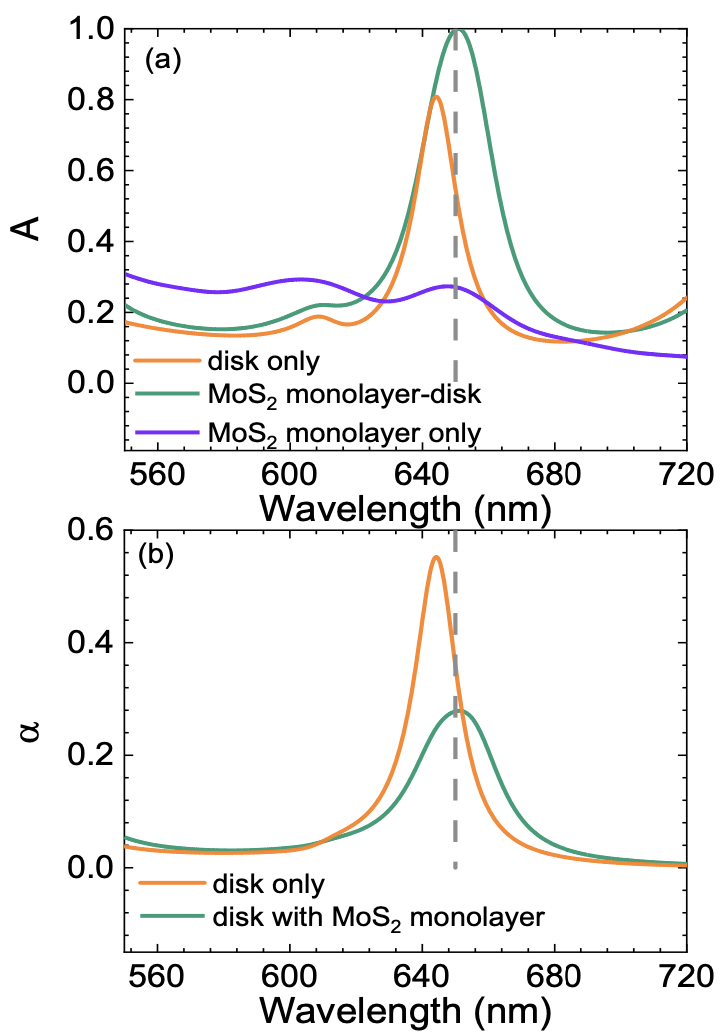}
\caption{\label{fig:PA} (a) Comparison of the absorbance, $\rm A$ of the structures on SiO$_2$/Au substrate for the disk radius $\rm r$ = 150 nm, the height, $\rm h$ = 45 nm and the spacer thickness of $\rm s$ = 100 nm, including the disk only, the monolayer MoS$_2$ only as well as the hybrid monolayer MoS$_2$-disk. (b) The absorbance $\rm \alpha$ of Si nanodisk only on SiO$_2$/Au substrate as well as the Si nanodisk in the hybrid monolayer MoS$_2$-disk on SiO$_2$/Au substrate. }
\end{figure}

It is interesting to explore further which multipolar modes of the Si nanodisk resonators drive the resonant absorbance enhancement. According to Ref. \cite{evlyukhin2016optical,zhang2020constructive}, the amplitude of decomposed multipolar modes contributing to the reflectance or transmittance coefficients are calculated by integrating the generated electric field $\rm \textbf{E}$ inside the nanodisk resonators for the array, including suspended in air or on substrates. The substrate effect on the reflectance spectra has been experimentally and theoretically explored in Ref. \cite{zhang2020constructive}, where the array of Si nanodisk resonators can be considered as a decoupled in-line optical element, where the modified generated electric field within the disk resonators due to the substrate need to be taken into account. It is clear from Fig.~\ref{fig:field} that the total electric dipole, including the electric dipole $\rm \textbf{P}$ and electric toroidal dipole contributions, $\rm \textbf{Te}$, as well as the magnetic quadrupole, $\rm \textbf{M}$ are dominant in the far-field reflectance or transmittance.  The magnetic dipole $\rm \textbf{m}$ and electric quadrupole $\rm \textbf{Q}$ are negligible.
Higher amplitude multipolar modes are seen for the disk resonator on SiO$_2$/Au substrate, which results from the higher generated electric field amplitude within the disk resonator compared with that in air or on SiO$_2$/Si substrate. Since the dominant multipolar modes, including the total electric dipole and magnetic quadrupole, have the same even parity in the forward and backward scattering plane \cite{evlyukhin2016optical,PhysRevB.103.195419}, the scattering profile does not depend on the incidence port, along $z$ or $-z$. Under the incident light along $z$, the forward scattering ($z$) is reflected by the gold mirror, which is incident on the disk resonators again, resulting in the other forward ($-z$) and backward scattering ($z$). Both scattering generated by the incident light as well as gold mirror reflection superimpose coherently. This corresponds well with the observation of near-identical spectral lineshapes for the decomposed multipolar modes of the disk resonators on SiO$_2$/Au substrate compared with 
those for the structure suspended in air in Fig.~\ref{fig:field} (a) and (c), only different with amplitudes, indicating that the gold plane plays the role of a second input port. Fabry-P$\rm \acute{e}$rot-type multiple reflections  are also present, however these can be treated as negligible \cite{Babicheva2017,zhang2020constructive}. 
Moreover, due to the constraint of power conservation in the cases of the perfect absorber and partial absorber, it is interesting to explore the effect of monolayer MoS$_2$ on the absorbance of disk resonators. From 
Fig.~\ref{fig:PA} (a), it can be seen that when the absorbance of the alone monolayer MoS$_2$ directly on the SiO$_2$/Au substrate is almost resonant with that disk resonators, the absorbance of the hybrid structure shows a red-shift due to the high imaginary permittivity of monolayer MoS$_2$. Furthermore, as can be seen in Fig.~\ref{fig:PA} (b), the absorbance of the Si nanodisks is reduced in the hybrid case and the resonance red-shifts due to the presence of monolayer MoS$_2$. 
This indicates that the incident energy redistributes in the hybrid structures, driven by relative imaginary permittivity of the disk and the monolayer MoS$_2$ as well as the modified electric field distributions within the disk and monolayer MoS$_2$ respectively compared with those for each separate element.

\section{conclusion}
In conclusion, the optical absorbance of monolayer
MoS$_2$ on an array of Si disk nanoresonators is explored experimentally and theoretically. 
Emission enhancement is experimentally observed for MoS$_2$ on the nanodisk array. This is due to increased absorbance in the MoS$_2$ monolayer, which also compensates for emission quenching due to the presence of the disk. 
The model explains the observed emission enhancement and is used to further explore increasing the magnitude of the absorbance of monolayer MoS$_2$ using Si nanodisk resonators.
With the aid of the gold mirror, a perfect absorbing structure is proposed, where the incident power is totally absorbed by the hybrid monolayer MoS$_2$-Si disk resonators. The dominant total electric dipole including toroidal dipole contributions as well as magnetic quadrupole drive the resonant absorbance enhancement. By analyzing the absorbance of the disk resonator and monolayer
MoS$_2$, respectively, it is seen that the full electromagnetic power is redistributed within the hybrid structures compared with those for the separate components. The absorbance of monolayer MoS$_2$ is enhanced while the absorbance of disk resonators is reduced. Moreover, compared to a monolayer
MoS$_2$ suspended in air, where the absorbance value of the A exciton peak value is smaller than 10\%, a value of more than 50\% is achieved in the perfect absorbing structures.

\begin{acknowledgments}
We wish to acknowledge the support of Science Foundation Ireland (SFI) under Grant Numbers 16/IA/4550 and 17/NSFC/4918.
\end{acknowledgments}

\nocite{*}

\bibliography{apssamp.bib}

\renewcommand\thefigure{S\arabic{figure}}
\textbf{Supplementary Material}
\subsection{Reflectance and Photoluminescence Measurement}

The nanodisk resonators array (50$\mu$m $\times$ 50$\mu$m) is illuminated by a white light beam and the reflected signal is collected by a 5$\times$, numerical aperture, NA = 0.12 lens, coupled to a CCD camera and Andor spectrometer. The incident angle is roughly 7$^0$ and incident beam spot is 6 $\mu$m. The reflectance spectra are normalized by the signal reflected from a silver mirror. The photoluminescence (PL) spectra are collected using the same setup with the laser excitation at $\lambda_{ex}$ = 405 nm and $\lambda_{ex}$ = 532 nm, respectively. The PL signal is collected using the same 5$\times$ microscope objective lens with NA = 0.12.

\subsection{Raman Measurement of Monolayer MoS2} 
Measurements for the PL and Raman maps were performed using a WITec Alpha 300R tool. The excitation laser was at 532 nm, operating at a power of $\sim$250 $ \mu W$. A 100$\times$ objective (NA = 0.95) was used. The Raman spectra/mapping were measured using a grating of 1800 g/mm, and the PL mapping were measured using a grating of 600 g/mm. The Raman spectra can be seen in Fig.~\ref{raman}. The in-plane ($ E_{2g}^{1}$) and out-of-plane ($ A_{1g}$) Raman modes are separated by $\Delta f \approx$ 18.2 $ cm^{-1}$ without disk resonator and $\Delta f \approx$ 19.2 $ cm^{-1}$ with disk resonator. 

These measurements confirm that the MoS2 directly on SiO$_2$/Si substrate and on Si disk/SiO$_2$/Si substrate is monolayer \cite{Lee2010,Li2012} and that the disk resonator has a very small effect on the frequency of the Raman modes. This suggests that the change of substrate does not introduce significant strain or have a significant impact on the doping level.

\begin{figure}[htbp]
\setcounter{figure}{0}
\includegraphics[width=0.9\linewidth]{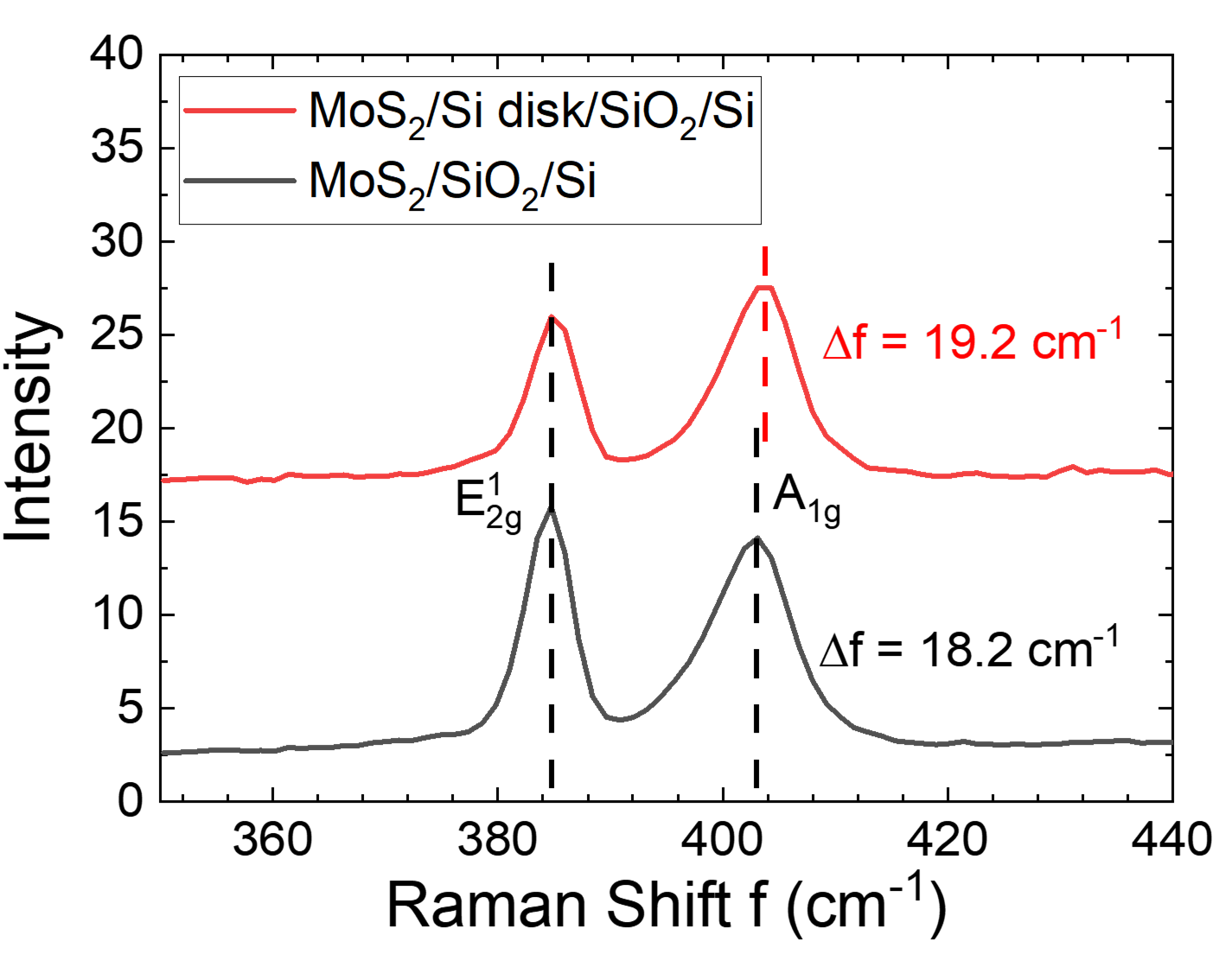}
\caption{\label{raman} Raman spectra of the grown MoS$_2$ on SiO$_2$/Si substrate and that on Si disk/SiO$_2$/Si substrate respectively: the in-plane ($ E_{2g}^{1}$) and out-of-plane ($ A_{1g}$) Raman modes are separated by $\Delta f=$ 18.2$\pm$0.5 $ cm^{-1}$ without disk resonator and $\Delta f=$ 19.2$\pm$0.5 $ cm^{-1}$ with disk resonator.}
\end{figure}

\subsection{Reflectance: Measurement and Simulation Comparison}
\begin{figure}[htbp]
\includegraphics[width=0.85\linewidth]{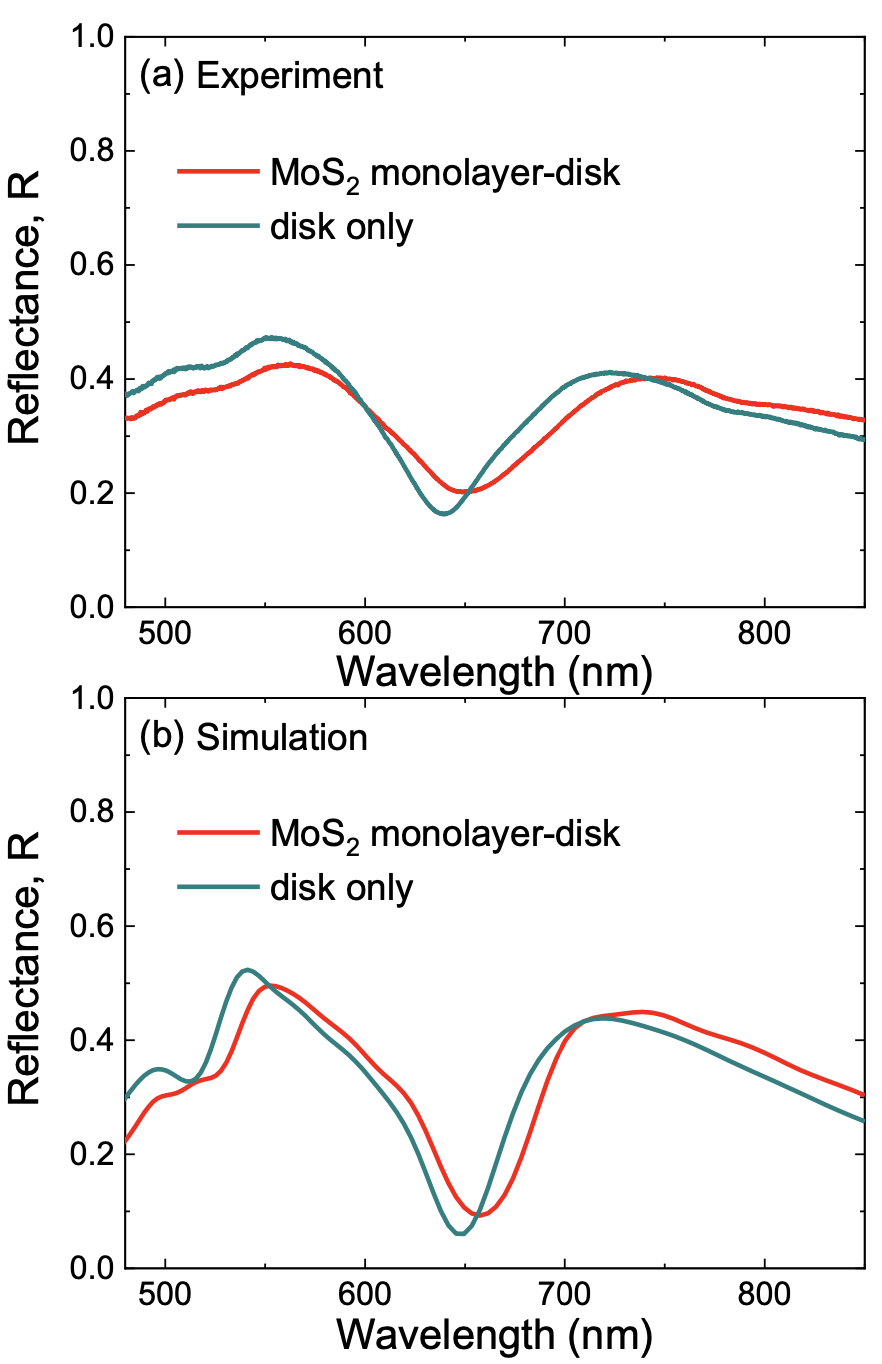}
\caption{\label{refR} The experimentally measured (a) and numerically simulated (b) reflectance, R spectra of the disk only and the hybrid MoS$_2$ monolayer-disk structures on SiO$_2$/Si substrate. The disk radius is $r$ = 150 nm, height $h$ = 45 nm, the SiO$_2$ layer thickness, $s$ = 140 nm.  }
\end{figure}

The measured reflectance spectra can be seen in Fig.~\ref{refR}, where the numerically simulated curves by Lumerical finite-difference-time-domain (FDTD) are also shown, which correspond well with the experimental measurement. 

\subsection{Theoretical Quantification of Photoluminescence Enhancement}

A resonator may change the PL intensity of an emitter, which is determined by the emitter's absorbance (excitation rate) as well as its quantum efficiency (Purcell factor). To quantify the The PL intensity changes due to the Si nanodisk resonators, $\eta$ is defined as the ratio of the integrated PL intensity of monolayer MoS$_2$ on top of Si nanodisk/SiO$_2$/Si substrate, $ I_{on-disk}$, to that placed directly on top of SiO$_2$/Si substrate, $ I_{off-disk}$, or

\begin{equation}
\label{PLcal}
   \eta=\frac{I_{on-disk}}{I_{off-disk}}= \frac{\alpha(\lambda_{pump})}{{\alpha}^{0}(\lambda_{pump})}\cdot\frac{q(\lambda_{em})}{q^0(\lambda_{em})} 
\end{equation}
where $ \alpha $ ($ \alpha^0 $) refers to the absorbance rate on disk (off disk) and $ q$ ($ q^0$) refer to the quantum efficiency on disk (off disk) respectively. Both of the absorbance rate and quantum efficiency can be calculated by FDTD. The wavelength-dependent absorbance spectra of monolayer MoS2 on disk or off disk is semi-analytically calculated by Eq.2 in the manuscript, or
\begin{equation}
\label{Eq:abspower}
\rm\alpha=\frac{P_{abs}}{P_{inc}}=\frac{k_0}{S_0|E_0|^2}\iiint_VIm[\epsilon (\textbf{r})]|\textbf{E}(\textbf{r})|^2 dV
\end{equation}
$| \textbf{E}(\textbf{r})|$ is the amplitude of the electric field within the layer of monolayer MoS$_2$, which can be directly obtained by the FDTD's electric field monitor placed at the position of the monolayer MoS$_2$.  The integration is performed in one unit cell over the volume $ V$ and $ V = p_x\times p_y\times t_0$, where $t_0$ denotes the thickness of the MoS$_2$ monolayer. From the calculated absorbance spectra of MoS$_2$ monolayer on disk and off disk respectively, the effect of Si nanodisk resonators can be clearly seen in Fig. 2 (d) upper panel in the main manuscript.  

Furthermore, the quantum efficiency $ {q (\lambda_{em})}$ and $ {q^0 (\lambda_{em})}$ is also calculated directly using FDTD software. The monolayer MoS$_2$ can be treated as a point dipole emitter which is averaged over the entire array. The dipole is oriented parallel to the substrate surface and is positioned over 11 different positions relative to the nanodisk resonator within one unit cell, with the top view ($x-y$) and side view ($x-z$) of the simulation setup seen in Fig. S3. The simulation is performed with the dipole emission at the monolayer MoS$_2$'s emission wavelength, at 673 nm and is averaged over 11 positions with a value of 0.74 obtained.

\begin{figure}[htbp]
\includegraphics[width=0.99\linewidth]{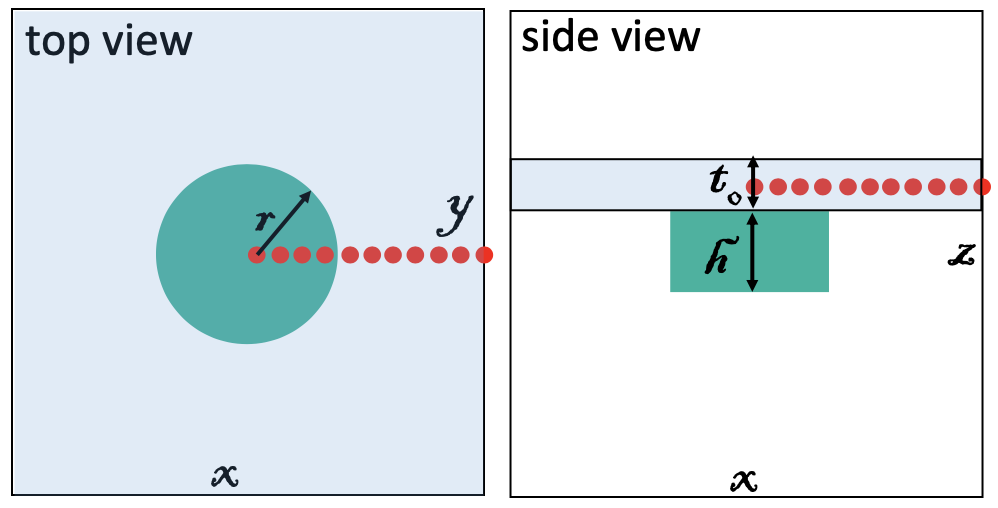}
\caption{\label{QEsim} The quantum efficiency simulation set up: the red dot denotes the placed single dipole emitter, which is positioned in the middle of the monolayer MoS$_2$ (illustrated in the shaded area in the side view), with 11 different positions relative to the nanodisk resonator. $ r$ denotes the radius of the Si nanodisk, $ h$ denotes the height of the Si nanodisk and $ t_0$ denotes the thickness of the monolayer MoS$_2$.}
\end{figure}

\end{document}